\documentclass[compsoc]{IEEEtran}
\usepackage[utf8]{inputenc}

\usepackage{url}
\usepackage{graphicx}
\usepackage[table,xcdraw]{xcolor}

\usepackage{color}
\usepackage{tikz}
\usepackage{qtree}
\usepackage{times}
\usepackage{epsfig}
\usepackage{graphicx}
\usepackage{epstopdf}
\usepackage{algorithm}
\usepackage{algorithmic}
\usepackage{amsmath}
\usepackage{amssymb}
\usepackage{amsxtra}
\usepackage{multirow}
\usepackage{mathtools}
\usepackage{cleveref}
\usepackage{url}
\usepackage{subfigure}
\usepackage{subfigure}
\usepackage{amsthm}
\usepackage{adjustbox}

\newtheorem*{proof*}{Proof}

\ifCLASSOPTIONcompsoc
  \usepackage[nocompress]{cite}
\else
  \usepackage{cite}
\fi

\begin{document}
%
\title{
IoT Supply Chain Security: Overview, Challenges, and the Road Ahead
}

\author{\IEEEauthorblockN{Muhammad Junaid Farooq and Quanyan Zhu\\}
\IEEEauthorblockA{Tandon School of Engineering,
New York University,
Brooklyn, NY 11201\\
Emails: \{mjf514,qz494\}@nyu.edu}\vspace{-0.2in}
}

\maketitle
\vspace{-0.2in}

\begin{abstract}

Supply chain is emerging as the next frontier of threats in the rapidly evolving IoT ecosystem. It is fundamentally more complex compared to traditional ICT systems. We analyze supply chain risks in IoT systems and their unique aspects, discuss research challenges in supply chain security, and identify future research directions.

\end{abstract}

%
\IEEEpeerreviewmaketitle

\section{Introduction}

The Internet of Things (IoT) is being used as a key enabling technology to secure the supply chain of several industries by tracking of assets, raw materials, and supplies. However, the supply chain security of the IoT itself is generally overlooked. The IoT is an interconnection of smart devices and components that come together to provide situational awareness and automated operation of electronic systems.  
It is not a standalone system obtained from a single supplier or manufacturer, having propriety hardware and software. Instead, it is composed of various different  interconnected components that may be designed, manufactured, and operated by different entities located in different parts of the world. 

\begin{figure}[t]
    \centering
    \includegraphics[width=2.5in]{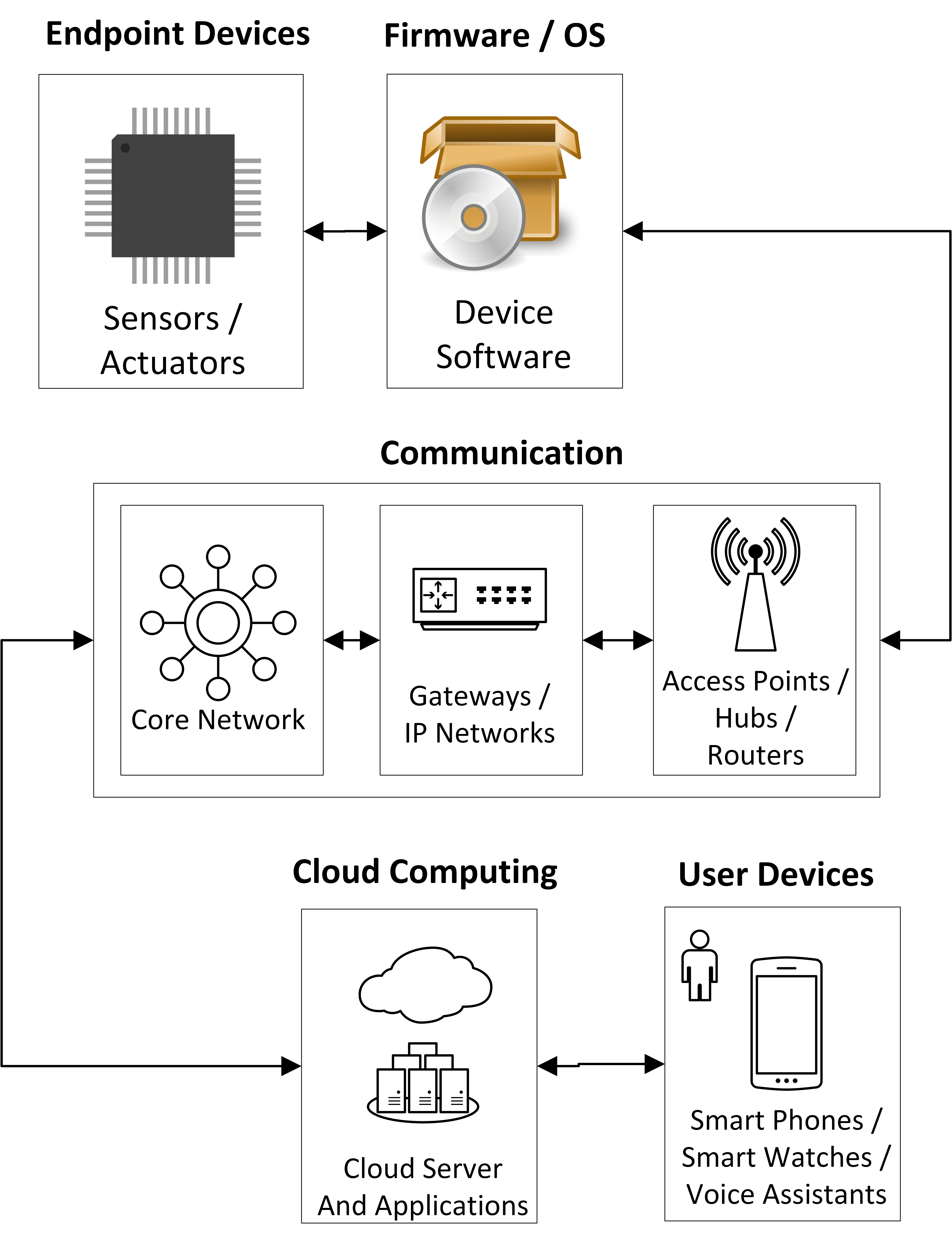}
    \caption{IoT Technology Stack.}
    \label{Fig:Stack}
\end{figure}

A generic illustration of the various components along the IoT technology stack and their interconnection is provided in Figure~\ref{Fig:Stack}. In essence, there are several actors involved in setting up the IoT ecosystem that include sensing/actuating device manufacturers, firmware developers, radio access network service providers, cloud service providers, mobile app developers, and end users. The endpoint devices are made of embedded hardware that interact with the physical environment and are driven by software processes referred to as firmware or operating system. These make use of communication infrastructure, which is composed of access points, gateways, and core IP networks to connect to cloud servers, that in turn host applications and services, which are operated by users via computing devices such as smart phones, smart watches and voice assistants, etc. 
More and more systems are becoming intelligent and autonomous with the emergence of IoT devices and enhanced ICT infrastructure. However, the integration of multiple devices and components that are designed and manufactured by different entities makes the system extremely vulnerable to cyber-physical attacks.

The IoT ecosystem faces serious security threats from traditional attackers due to factors such as low cost, inherent inter-operability, and rapid product development life cycle. Recent large scale cyber attacks such as  \emph{Mirai}~\cite{mirai} botnet and \emph{Stuxnet}~\cite{stuxnet} have exploited some of the vulnerabilities in IoT systems. Apart from other traditional cyber and physical threats to the IoT ecosystem, the supply chain is emerging as a new source of potential threats. It emanates from the fact that the IoT systems are often deployed in a decentralized manner where managers acquire and deploy equipment needed for improving the efficiency of their business lines. It is generally done without investigating who the suppliers actually are and what risks the suppliers bring to the overall system. Small enterprises, in general, do not have the necessary resources to manage or even assess their risks by increasing automation via the IoT~\cite{best_practices}. Therefore, besides tackling traditional cyber-physical threats to the IoT, steps need to be taken to ward off threats at the supply chain front as well. This article provides an analysis of some of these threats, research challenges, and potential future directions. We restrict our focus on the supply chain dimension of attacks and risks in the IoT arena.



\section{IoT Supply Chain Risk Landscape: Hard to Observe and Hard to Control}

Supply chain risk has long been a matter of great concern for businesses and corporations. In fact, supply chain risk management (SCRM) is a standard functional area across many industries such as consumer goods, food, industrial products, etc., and is considered to be a vital component in securing the revenues and profitability of enterprises. 
Security of information and communications technology (ICT) equipment has been an area of immense focus in recent times. 
since more advanced and sophisticated methods have emerged to attack IT/OT systems.
Cyber-physical attacks on these systems may result in significant monetary and non-monetary losses. To counter threats from such attacks, the U.S. National Institute of Standards and Technology (NIST) has prepared a comprehensive list of best practices for SCRM in traditional ICT systems~\cite{nist_report}.

The development and growth of the IoT is further enhancing the security concerns. Although the flexibility of communication and interaction between devices results in tremendous benefits, however it also opens doors for attackers and malicious actors to sabotage the system. With the emergence of vendor based attacks and the involvement of global players, there are rising concern on the security of the IoT supply chain. 
The IoT is a special class of ICT systems and is evolving rapidly. The interconnection of systems and devices enables a much richer attack surface as opposed to traditional ICT systems. Moreover, the supply chain of the IoT is extremely complex, globally distributed, and highly inter-connected. In addition, the IoT is still a grossly unregulated technology in terms of security standards unlike food, where the risks are better understood. It is mainly because the ecosystem is highly diverse and the consequences of attacks are relatively unknown. In certain industries such as food and medicine, there are agencies that regulate the safety standards. It's because the risk assessment has been done by testing the products repeatedly on subjects and evaluating the results. 
However, in the IoT ecosystem, there are limitless functionalities as well as possibilities of malfunction and malicious activity. Hence, determining the possible attacks and enumerating the consequences becomes extremely challenging.

In summary, the risk landscape of the IoT supply chain is extremely diverse. The suppliers may ember backdoor channels un devices, inject viruses, provide faulty chips, or load with malicious software. These are some of the possibilities that the IoT systems can be attacked. The alarming concern is that these IoT systems are set to control national critical infrastructure resources as well as improving battlefield effectiveness. The supply chain risks are hard to observe and hard to control. The risk propagates from one device to the other and gets amplified as the IoT ecosystem becomes more complex. It is not straightforward to determine where to regulate the entire system.



\section{Dissecting Supply Chain Links in IoT}
\begin{figure}[t]
	\centering
	\includegraphics[width=3in]{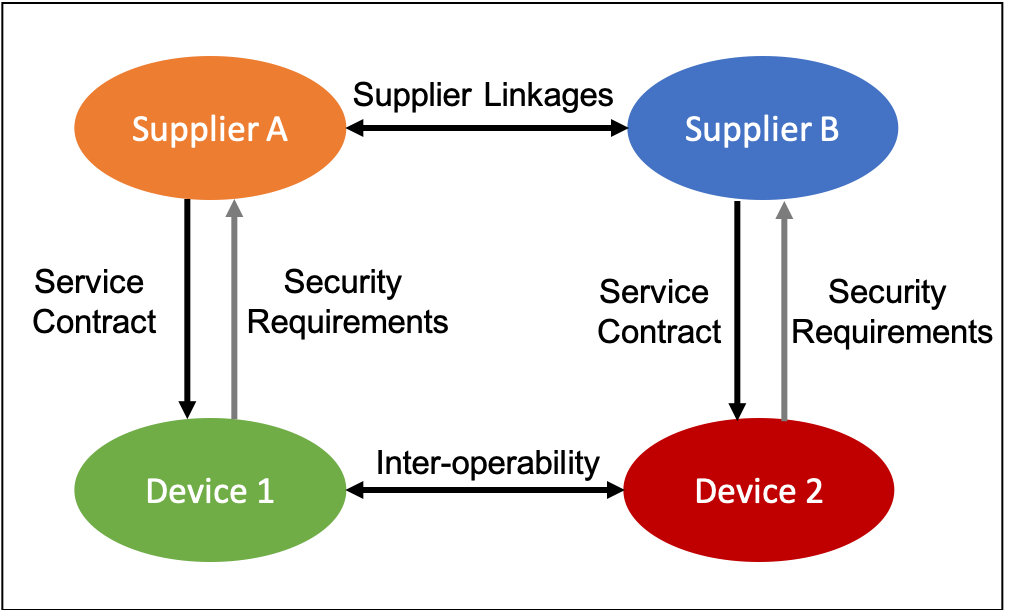}
	\caption{Key interactions between different players in the supply chain ecosystem of the IoT.}
	\label{Fig:Interactions}
\end{figure}

There is a delicate interplay between suppliers and devices in an IoT ecosystem. To illustrate the different type of interactions that may be present between suppliers and devices, we provide an example in Fig.~\ref{Fig:Interactions} where there are two devices obtained from two different suppliers. While the supply chain can be constructed several levels deep due to individual components in devices being manufactured by different entities, however, for the sake of simplicity, the immediate supplier of standalone devices is considered. In such a scenario, the following different interactions between the supply chain actors might be present.
\begin{itemize}
    \item \textbf{Device-Supplier Interactions:} This is a typical buyer supplier interaction. The devices are procured from the suppliers and have service contracts including maintenance, upgrades, security patches, etc. The devices have security and support requirements that need to be met under the agreements. 
    \item \textbf{Supplier-Supplier Interactions:} Suppliers may have different front-end companies but common connections at the back end. This is typically common in the tech world where corporations have mergers and takeovers. Different suppliers may be owned by a common entity having more control over the supply chain of the IoT network. A nexus of supply chain actors may result in the possibility of coordinated attacks using backdoor channels and other forms of advanced persistent threats.
    \item \textbf{Device-Device Interactions:} These interactions are present due to the inter-connectivity of the IoT devices to provide desired functionality. These interactions are significant since they allow supply chain risks from one device to transfer to the other independently of its own supply chain.
\end{itemize}


\begin{figure}
	\centering
	\includegraphics[width=3in]{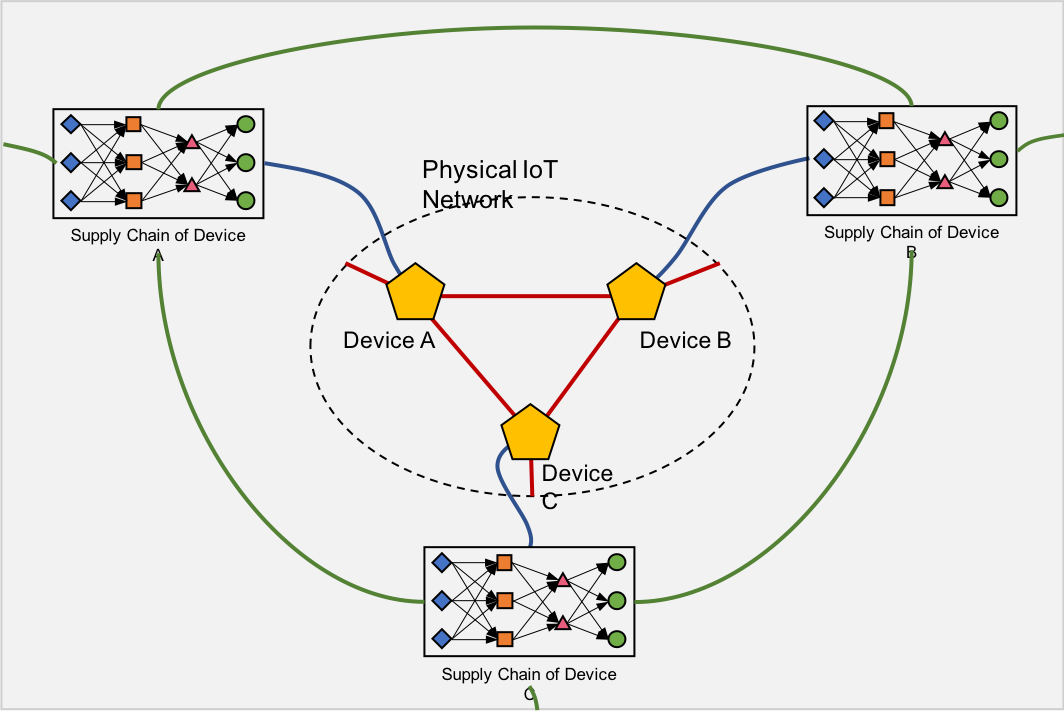}
	\caption{Mapping of IoT and the underlying supply chain networks.}
	\label{Fig:Mapping}
\end{figure}

A more detailed illustration of the supply chain interaction with the physical IoT network is provided in Fig.~\ref{Fig:Mapping}. There is a component graph that defines the underlying connectivity of devices that make up the IoT ecosystem. Each component has its independent supply chain. However, the supply chains of devices may be linked not only via external affiliations but also via physical connectivity of devices in the IoT network. It implies that the risk reciprocates among devices in the network. In other words, my risk becomes your risk and your risk becomes my risk. This makes the analysis of supply chain risks in IoT systems extremely convoluted.

\section{IoT Risk Implications and Consequences}

The impact of risk in IoT is critically important to analyze since it deals with the physical world and any attack or malicious activity may result in significant consequences such as physical damages, operational disruption, danger to human safety, etc. For instance, a malfunctioning of heating or cooling systems may result is sudden power surges resulting in breakdowns. It may cause a significant loss of revenue as well as damage to the power system. Therefore, there are implications and consequences of the risk that can be categorized as follows:
\begin{itemize}
    \item \emph{Monetary Implications:} The risk inevitably translates to monetary impact since any disruption or damage to infrastructure would lead to loss of revenue and/or safety hazards. Therefore, the financial impact is important to take into account while selecting the supply chain of IoT network.
    \item \emph{Legal Implications:} In the occurrence of a large scale cyber incident, the liability and responsibility needs to be determined. It is important to determine to what extent are the supply chain actors liable for security breaches and what actions can be taken against them. Therefore, there is a need to clearly map out the liability network.
    \item \emph{Policy Implications:} Risk determines a lot of policies that must be followed by the IoT ecosystem. Patching and upgrade policies are determined by how risky the system is. Cyber-insurance policies and premiums will also depend on how much risk is present in the system. While supply chain is only a part of the total risk, however, it plays a crucial role is determining the cyber risk of the overall system since the same functionalities may be offered by less trustworthy suppliers. 
\end{itemize}

It is important for stakeholders in the supply chain to be aware of the implications so they can decide whether they want to be part of a supply chain that may render them legally or financially liable.

\section{Unique Aspects of IoT Security}

The IoT itself and its security is drastically different than traditional ICT systems. Firstly, there are many different players participating in an unregulated system. Then, the devices are highly inter-operable allowing for limitless possibilities for applications. In fact, it is up to the individual users to build their own desired functionalities and enforce policies on the system. Unlike the Internet, there is no standard protocol stack for the IoT ecosystem. This makes it difficult to embed security into the protocols. The inherent inter-operability in the IoT creates security challenges and vulnerabilities. To this end, the role of the supply chain in IoT is also completely different. While existing wisdom in SCRM for ICT does act as a useful guideline, it may not be sufficient to tackle the more complex nature of IoT networks and the associated supply chain~\cite{implications}. A summary of the key differences in IoT systems and their security as compared to conventional ICT systems is provided below:
\begin{itemize}
    \item \emph{Interaction with the physical world:} The IoT devices interact with the physical world with the aid of actuation capabilities as opposed to conventional mobile and computing systems. It results in completely different consequences compared to ICT systems since it may endanger human safety, damage equipment, or cause operational disruptions.
    \item \emph{Limited access control and management:} The IoT devices are often low powered with limited computational capabilities. The complete access and management functionalities may not be built into these devices.
    \item \emph{Different cyber-security standards:} The security and privacy requirements for IoT device operation may be completely different from conventional ICT systems due to different authentication and access mechanisms.
    \item \emph{Heterogeneous ownership and de-centralized control:} There is no network administrator that has control over the IoT device configurations. Unlike routers and IP networks, the network administrator may not even have a complete view of all the connected devices in the network. 
\end{itemize}



\section{Research Challenges}
Since, the IoT is inherently a de-centralized system, it is difficult to exert control over the entire supply chain. However, the challenges go much beyond the regulation of the supply chain~\cite{mapping_security_challenge}. It is important to study and analyze the threat ecosystem in the IoT landscape. It implies that the potential sources of attack be identified and their potential implications studied in terms of the functionality and/or damage caused to the overall system. In this aspect, some challenges are related to technical aspects of IoT, while others emanate from the logistics and analysis or decision-making standpoint~\cite{nist_report_2}.
Some of the key logistical challenges are as follows:
\begin{itemize}
    \item \emph{Lack of control over upstream supply chain:} There is no control over upstream supply chain from a device owners point of view. 
    In other words, the buyers do not have complete information about the cyber-physical supply chain of the products.
    \item \emph{Disclosure of supply chain information:} Not all suppliers are ready to clearly articulate their cyber security practices and disclose their supply chain information. Some of it is obvious due to privacy reasons and competitor-sensitive information.
    \item \emph{Awareness of vulnerabilities:} The suppliers of IoT equipment may not be fully aware of all the possible vulnerabilities in their products. This makes it harder to determine the possible attack paths and analyze risk.
    \item \emph{Centralized database of vulnerabilities:} There is no centralized database of known vulnerabilities and attacks that can serve as a guideline to identify risks and possible attacks.
    \item \emph{Heterogeneous supply chain management practices:} The management practices for supply chain to mitigate associated security risks are diverse and depend on the industry. heterogeneity across application sectors.
\end{itemize}

Apart from the logistical constraints, there are also technical challenges in managing security of the IoT devices can be expressed as follows:
\begin{itemize}
    \item \emph{Lack of management controls:} centralized network management may not be available for the IoT. There is a need for developing management platforms to provide more control to the administrator over the IoT infrastructure.
    \item \emph{Inflexible hardware:} IoT device hardware may not be serviceable, meaning it cannot be repaired, customized, or inspected internally
    \item \emph{Heterogeneous ownership:} The devices are owned and operated by separate entities resulting in less control over policy implementation.
\end{itemize}

Finally, some of the decision-making and policy questions that need to be addressed are as follows:
\begin{itemize}
    \item \emph{Risk informed procurement and deployment:} The decisions to procure and deploy IoT network devices needs to be done in a risk-informed manner to allow for cost-benefit analysis.
    \item \emph{Contingency planning:} The IoT network requires arrangement of contingencies since suppliers may end security updates or discontinue support for the equipment.
    \item \emph{Risk-Conscious supplier contracts:} Contracts for installation services should include risk as as essential factor to enable a more secure infrastructure.
\end{itemize}

Tackling these challenges by finding out novel ways to counter them is important for the research community and policy makers. These are some of the opportunities for researchers and technologists to come up with ways to counter these different types of challenges that will pave the way for securing the IoT ecosystem from supply chain threats.

\section{The Way Forward}
In a complex system of systems setting with many supply chain actors and their convoluted interactions, it is an extremely challenging problem to mitigate and control supply chain oriented risks. Moving forward, the way to tackle the problem is to first fully understand the ecosystem from a supply chain viewpoint and then take appropriate measures to control the risks. 
In light of the highlighted challenges, there are two possible approaches that can be followed to tackle the problem. The first is the top-down approach, which is more centralized, while the second one is the bottom-up approach, which takes a de-centralized viewpoint.




\subsection{Top-Down approach to managing risk}
\begin{figure}[h]
    \centering
    \includegraphics[width=1.7in]{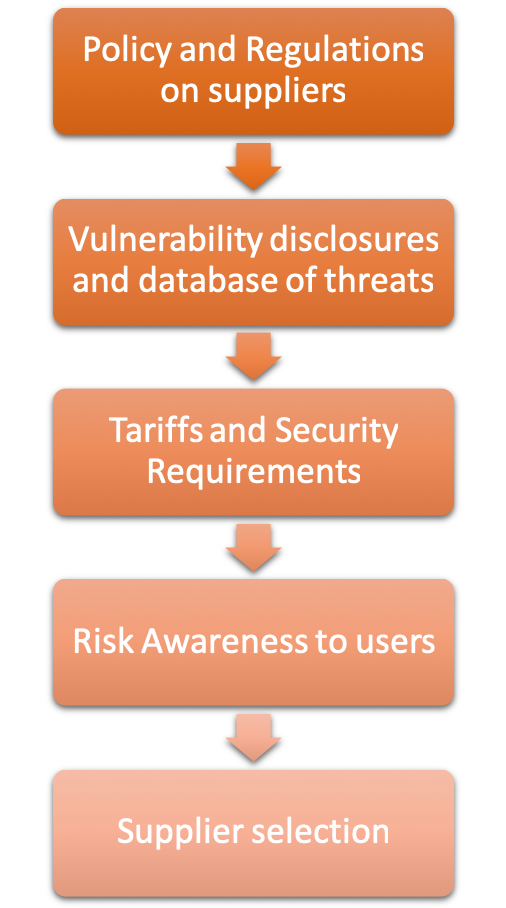}
    \caption{Top-down approach for managing supply chain risks.}
    \label{Fig:top_down}
\end{figure}

Governments and policy makers are unable to micro manage and control individual users of technology to adopt certain practices particularly in the technology supply chain front. A top-down approach uses a regulatory view of controlling supply risks in the IoT ecosystem. At the outset, policies and restrictions can be imposed on certain supply chain actors. For instance, certain suppliers of equipment may be banned for use in an industry due to detected malicious practices or excessive testing and standards may be enforced on certain suppliers based on their trust and reliability levels. Furthermore, they can be mandated to have compulsory disclosure of vulnerabilities to form a centralized database of threats. These will eventually lead to imposition of tariffs and security requirements on the suppliers. Once the policies are in place, the hope is that the managers and users of IoT technology will be aware of the risks they import by procurement from certain suppliers. This will ultimately result in risk-aware decisions by the users of technology based on other considerations such as cost and functionality. In essence, using a top-down approach, policy drives the underlying technology and supply chain actions. The hope is that centralized awareness and decision making may have a trickle down effect to secure the IoT ecosystem from the supply chain threats. Eventually, it might lead to a the development of secure supply chain architectures for IoT~\cite{architecture} ecosystems. An illustration of the main stages in the top-down approach is provided in Fig.~\ref{Fig:top_down}.



\subsection{Bottom-Up approach to managing risk}

\begin{figure}[h]
    \centering
    \includegraphics[width=2in]{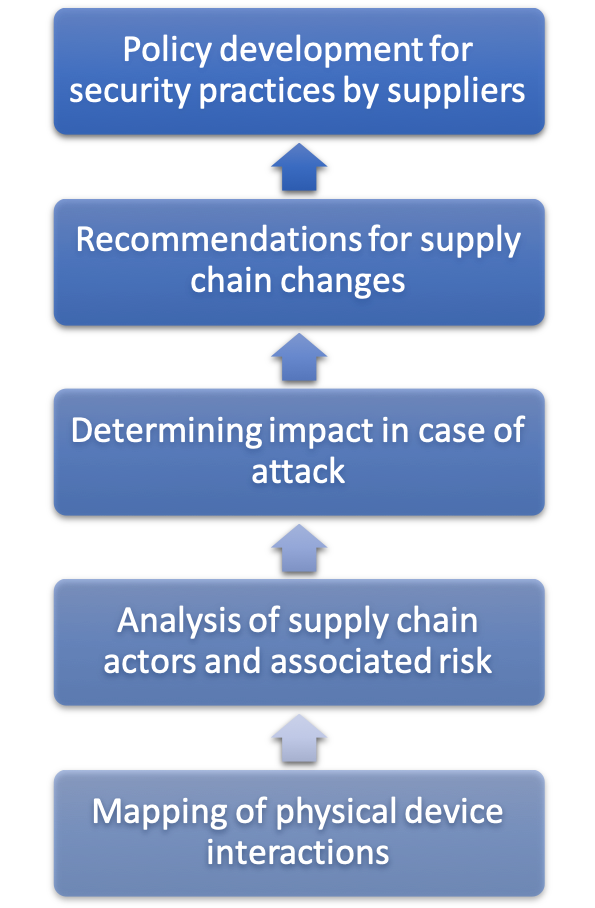}
    \caption{Bottom-up approach for managing supply chain risks.}
    \label{Fig:bottom-up}
\end{figure}

The bottom up approach uses a totally different view from the top-down approach. It aims to map out the view first and then lead to the development of policies to control the risks. The first step is to study and analyze the threat landscape, i.e., sources of attack and potential impacts in terms of functionality and anticipated loss/damage caused. This enables the formation of a comprehensive view of threats and vulnerabilities that are both adversarial and non-adversarial. 
Once the view has been mapped out, there needs to be a more holistic and integrated measurement of risk. Compound metrics for analyzing the risk as well as the impact are needed. The risk is generally measured as the impact times the likelihood. While the likelihood can be determined using attack trees, the impact needs to be studied more carefully by examining the inter-dependencies and information flows. Then, the goal is to develop mitigation strategies. New infrastructure can then be developed such as a centralized management platform that is in control of the administrators to have a network wide view of the IoT ecosystem and the supply chain actors involved. While platforms may not be required for individual home users of IoT devices, however, enterprises may need to have a comprehensive tool that allows them to have a clear map of their deployments and the associated risks and propagation. Once this is done, then policies and best practices can be developed for wider dissemination and enforcement. An example of such policy guidelines is the strategic principles that have been proposed by the U.S. Department of Homeland Security for securing the IoT~\cite{dhs_strategic_principles}. Consequently, road maps for implementation can be developed by individual industries according to their requirements~\cite{utc_management}.
The bottom up approach can be summarized in the flow shown in Fig.~\ref{Fig:bottom-up}. In essence, the technology and risk assessment drives the development of policy and regulations.




\section{Conclusion}

The supply chain risk management in IoT is a highly convoluted problem to solve. Going forward, we envision that advances will be made at several different fronts. The first direction is the accurate mapping of the threat landscape in terms of the supply chain. Then, a comprehensive risk assessment and impact analysis needs to be done. Finally, mitigation strategies are needed to act as a guideline for establishing best practices. Effective solutions to ensure the supply chain security of the IoT requires a long trajectory of development. There is a need for active public-private engagement to come up with concrete solutions. Joint policy and technical solutions are needed to counter the risks. Technology should inform policy and policy should regulate technology. Continuous assessment of risks presented by existing suppliers and response strategies is required for an effective defense against the emerging supply chain threats in IoT systems.





%



\bibliographystyle{IEEEtran}
\bibliography{references}

\begin{IEEEbiographynophoto}{\textbf{\\Muhammad Junaid Farooq}} (S'15) received the B.S. degree in electrical engineering from the School of Electrical Engineering and Computer Science (SEECS), National University of Sciences and Technology (NUST), Islamabad, Pakistan, the M.S. degree in electrical engineering from the King Abdullah University of Science and Technology (KAUST), Thuwal, Saudi Arabia, in 2013 and 2015, respectively. Then, he was a Research Assistant with the Qatar Mobility Innovations Center (QMIC), Qatar Science and Technology Park (QSTP), Doha, Qatar. Currently, he is a PhD student at the Tandon School of Engineering, New York University (NYU), Brooklyn, New York. His research interests include modeling, analysis and optimization of wireless communication systems, cyber-physical systems, and the Internet of things. He is a recipient of the President's Gold Medal for academic excellence from NUST, the Ernst Weber Fellowship Award for graduate studies and the Athanasios Papoulis Award for teaching excellence from the department of Electrical \& Computer Engineering (ECE) at NYU Tandon School of Engineering.
\end{IEEEbiographynophoto}

\begin{IEEEbiographynophoto}{\textbf{\\Quanyan Zhu}} (S'04, M'12) received B. Eng. in Honors Electrical Engineering from McGill University in 2006, M.A.Sc. from University of Toronto in 2008, and Ph.D. from the University of Illinois at Urbana-Champaign (UIUC) in 2013. After stints at Princeton University, he is currently an assistant professor at the Department of Electrical and Computer Engineering, New York University. He is a recipient of many awards including NSERC Canada Graduate Scholarship (CGS), Mavis Future Faculty Fellowships, and NSERC Postdoctoral Fellowship. He spearheaded and chaired INFOCOM Workshop on Communications and Control on Smart Energy Systems (CCSES), and Midwest Workshop on Control and Game Theory (WCGT). His current research interests include resilient and secure interdependent critical infrastructures, energy systems, cyber-physical systems, and cyber-enabled sustainability. He is a recipient of best paper awards at 5th International Conference on Resilient Control Systems, and 18th International Conference on Information Fusion. He has served as the general chair of the 7th Conference on Decision and Game Theory for Security (GameSec) in 2016 and International Conference on NETwork Games, COntrol and OPtimisation (NETGCOOP) in 2018.
\end{IEEEbiographynophoto}

\end{document}